\documentclass{ws-procs975x65}

\begin{document}

\title{LATE-TIME EXPANSION IN THE SEMICLASSICAL THEORY OF
  HAWKING RADIATION} 

\author{PIETRO MENOTTI$^*$}

\address{Department of Physics, University of Pisa,\\
Pisa, 56127, Italy\\
$^*$E-mail: menotti@df.unipi.it\\
www.df.unipi.it/$\sim$ menotti}

\begin{abstract}
 We give a detailed treatment of the back-reaction effects on the Hawking
  spectrum in the semiclassical approach to the Hawking radiation. 
We solve the exact system of non linear equations giving
  the action of the system, by a rigorously convergent  iterative
  procedure. The first two terms of such an expansion give the
  $O(\omega/M)$ correction to the Hawking spectrum.
\end{abstract}


\bodymatter

\section{Introduction}\label{aba:sec1}

The semiclassical treatment of the Hawking radiation was introduced by Kraus
and Wilczek \cite{KWI}.
The interest of the approach is to provide a method to compute the
back-reaction effect of the radiation on the black hole,
an effect which is completely ignored in the the external field
treatment of the phenomenon.
The main idea is to replace the free field modes of the
radiation by the semiclassical wave function of a shell of matter or
radiation which consistently propagates in the gravitational field generated
by the back hole and by the shell itself. The shell dynamics was studied in
detail in many papers (see \cite{KWI,FLW,FM,menottiphysrev}). 
In the original semiclassical treatment 
\cite{KWI,KVK} the spectrum of the
Hawking radiation is extracted through the standard Fourier analysis of the
regular modes. Later such a treatment was related to the tunneling
picture
\cite{PW};
such an approach gave also rise to several proposals and to controversy
\cite{chowdhury,akhmedov,zerbini1,zerbini2,pizzi,vanzo}.

We think that  the mode analysis is still the clearest 
and safest way to extract the results in the semiclassical approach. 

The present work \cite{menottiphysrev} is devoted to a detailed analysis of the
construction of the semiclassical modes and their time Fourier transform. The
action related to the modes which are regular on the horizon is defined
through mixed boundary conditions, i.e. a condition on the value of the
conjugate momentum at $t=0$ and a condition at time $t$ on the
coordinate $r$. 
The computation of the action as a function of $t$
corresponds to the solution of a system of two highly non linear
equations where the two unknown are the value $H$ of the Hamiltonian
and the shell position at time $t=0$, $r_0$ which also depends on the
mixed boundary conditions.  
In \cite{KWI} 
a truncated system of equations obtained by keeping only the most
singular terms in the exact equations was considered. Through a long chain
of approximations the authors reached for the effective temperature, due to
the back reaction effects, the value $1/(8\pi M (1-\omega/M))$. Later
Keski-Vakkuri and Kraus \cite{KVK} using a completely different method 
obtained for
such effective temperature the value $1/(8\pi M (1-\omega/2M))$.
Here we reconsider the problem along the lines of \cite{KWI}
treating the full exact system of equations. 
We show that such a system of equations is
equivalent to an other non linear equation which can be solved by a convergent
iterative procedure. The first two terms of the convergent
iterative procedure are sufficient to provide the leading spectrum of the
radiation and its back-reaction correction terms of order $\omega/M$
confirming the result of \cite{KVK}.

\section{Choice of gauge, the action and the equations of motion}

The rotationally invariant Painlev\'e-Gullstrand metric is given by
\begin{equation}\label{metric}
ds^2 =-N^2 dt^2+(dr+N^r dt)^2 +R^2 d\Omega^2
\end{equation}
where all quantities $N,N^r,R$ are functions only of $r$ and $t$.
One has still a gauge choice on $R$. In
presence of a shell of matter we shall use the ``outer gauge''
\cite{FM,menottiphysrev} which is defined by $R=r$ for $r\geq \hat
r$ where $\hat r$ denotes the shell position. At $r=\hat r$, 
$R$ is continuous as all the other functions appearing in (\ref{metric}), 
but its derivative is discontinuous. 
After solving the constraints the action takes the form in the
massless case \cite{KWI,FLW,FM,menottiphysrev} 
\begin{equation}\label{reducedaction3}
S = \int_{t_i}^{t_f} \big(p_{c}~ \dot{\hat r} - H)dt
~~~~{\rm where}~~~
p_c =\sqrt{2M \hat r}-\sqrt{2H \hat r}-\hat r \log\frac{\sqrt{\hat
r}-\sqrt{2H}}{\sqrt{\hat r}-\sqrt{2M}}.
\end{equation}    

At the semiclassical level the modes which are invariant under the Killing
vector $\frac{\partial}{\partial t}$ are simply given by

\begin{equation}
e^{iS/l_P^2}~~~~{\rm with}~~~~S = \int^{r_1} p_c dr - Ht +{\rm const}.
\end{equation}    
where $l_P^2= G\hbar$ is the square of the Planck length.
As it is well known such modes have the feature of being singular at
the horizon.
Instead the true vacuum should be described in term of modes which are regular
at the horizon 
for which an outgoing shell of matter has the following boundary 
conditions \cite{KWI}: 
i) at time $0$ the conjugate momentum is a given value $k$; 
ii) at time $t$ the shell position $r$ is a given value $r_1$. The expression
for such an action was already given by Kraus and Wilczek in \cite{KWI}.
With the two conditions $p_c(0) =k$ and $r(t) = r_1$ the  action is
\begin{equation}\label{KWaction}
S(r_1,t,k) = 
k r_0(r_1,t,k) +\int_0^{t}p_c \dot{r} dt' - 
H[r_1,t,k] t .
\end{equation}
$r_0$ denotes the value of $r$ at time $0$; also such a
quantity depends on the imposed boundary conditions. 

In the outer gauge \cite{KWI,FM,menottiphysrev} the equations of motion can be
integrated to
\begin{equation}\label{tequation}
t=
4H\log\frac{\sqrt{r_1}-\sqrt{2H}}{\sqrt{r_0}-\sqrt{2H}}+r_1-r_0+2\sqrt{2Hr_1} 
-2\sqrt{2Hr_0}.
\end{equation}
with the boundary condition at $t=0$
\begin{equation}\label{kequation}    
0<k =\sqrt{2M r_0}-\sqrt{2H r_0}- r_0 \log\frac{\sqrt{r_0}-\sqrt{2H}}
{\sqrt{r_0}-\sqrt{2M}}
\end{equation}    
where $2M<2H<r_0<r_1$. Eq.(\ref{kequation})
together with eq.(\ref{tequation}) should determine completely $H$ as
a function of $t$. 
However with the above mentioned mixed boundary condition in
general caustics arise \cite{menottiphysrev}, 
i.e. in general more that one trajectory in phase space
satisfies the mixed boundary conditions.
On the other hand we prove \cite{menottiphysrev}
that if the end point $r_1$, is less that a
critical value $r_c$, caustics to not occur.

\section{The late-time expansion}

The problem now is to solve the system (\ref{tequation},\ref{kequation}). 
After introducing the implicit time $T=e^{-\frac{t}{4H}}$ 
and using the notation $ h = \sqrt{2H} ,~~m = \sqrt{2M}$
it is possible to rewrite the system of equations
(\ref{tequation},\ref{kequation}) 
as \cite{menottiphysrev}
\begin{equation}\label{implicit}    
h = m+T f(T) \equiv m+g(T)
\end{equation}
with $g(0)=0,g'(T)>0$. It is then possible,
starting from $h_0=m$ to solve eq.(\ref{implicit}) iteratively, 
where the process converges rigorously at all times \cite{menottiphysrev}.
The first two terms give
for the semiclassical mode
\begin{equation}\label{actionexp}
e^{iS/l_P^2}
= e^{i [q(r_1) - Mt +4 M\sqrt{2M} \tau_1 + t \tau_1^2]/l_P^2}~~~{\rm
  with}~~~
\tau_1 = c(k,r_1)e^{-t/(4M)}.    
\end{equation}
Using the modes (\ref{actionexp}) to compute the Bogoliubov
coefficients one obtains 
\cite{menottiphysrev} for the flux of the Hawking radiation
\begin{equation}
F(\omega) d\omega =\frac{d\omega}{2\pi}\frac{1}{e^{8\pi
\frac{M\omega}{l_P^2}(1-\frac{\omega}{2M})}-1}~.
\end{equation}
in agreement with \cite{KVK}.


\begin{thebibliography}{9}

\bibitem{KWI} P. Kraus, F. Wilczek, Nucl.Phys. B433 (1995) 403,
  Nucl.Phys. B437 (1995) 231.

\bibitem{menottiphysrev} P. Menotti, Phys.Rev. D85 (2012)
084005-1; Class.Quant.Grav. 27:135008 (2010),
J.Phys.Conf.Ser. 222:012051 (2010).  


\bibitem{FLW} J. L. Friedman , J. Louko, S. Winters-Hilt, Phys.Rev. D56
(1997) 7674.

\bibitem{FM} F. Fiamberti, P. Menotti, Nucl.Phys. B794 (2008) 512.

\bibitem{KVK} E. Keski-Vakkuri, P. Kraus, Nucl.Phys. B491 (1997) 249.

\bibitem{PW} M. K. Parikh, F. Wilczek, Phys.Rev.Lett. 85 (2000) 5042.

\bibitem{chowdhury} B. Chowdhury, Pramana 70 (2008) 593, 
P. Mitra, Phys.Lett. B648 (2007) 240.

\bibitem{akhmedov} E. T. Akhmedov, V. Akhmedova, D. Singleton,
Phys.Lett. B642 (2006) 124; V. Akhmedova, T. Pilling, A. de Gill, D.
 Singleton, Phys.Lett. B666 (2008) 269.

\bibitem{zerbini1} S. A. Hayward , R. Di Criscienzo, L. Vanzo, M. Nadalini,
S. Zerbini, Class.Quant.Grav. 26:062001 (2009).

\bibitem{zerbini2} R. Di Criscienzo, S. A. Hayward, M. Nadalini,
L. Vanzo, S. Zerbini, e-Print: arXiv:0906.1725 [gr-qc]; S.A. Hayward, 
R. Di Criscienzo, M. Nadalini, L. Vanzo, S. Zerbini, AIP
Conf.Proc.1122:145 (2009), e-Print: arXiv:0812.2534 [gr-qc]; 
e-Print: arXiv:0909.2956 [gr-qc].

\bibitem{pizzi} M. Pizzi, arXiv:0904.4572 [gr-qc]; arXiv:0907.2020
[gr-qc]; arXiv:0909.3800 [gr-qc].

\bibitem{vanzo} L. Vanzo, G. Acquaviva, R. Di Criscienzo,Class.Quant.Grav. 
28 (2011) 183001. 

\end{thebibliography}
\end{document}